\documentclass{article}
\usepackage{graphicx}
\usepackage[varg]{txfonts}
\usepackage{xcolor}
\usepackage{verbatim}
\usepackage{textcomp, gensymb}
\usepackage{acronym}
\usepackage{natbib}
\usepackage[colorlinks=true,citecolor=blue]{hyperref} 
\usepackage{tablefootnote}
\usepackage{multirow}
\usepackage{tabularx,booktabs}
\usepackage{subcaption}
\usepackage{makecell}

\usepackage{authblk}
\usepackage[margin=1.5in]{geometry}
\usepackage{xspace}
\usepackage[labelfont=bf]{caption}

\DeclareCaptionFormat{cont}{#1 (cont.)#2#3\par}

\newcolumntype{C}{>{\centering\arraybackslash}X}
\newcolumntype{L}{>{\raggedright\arraybackslash}X}
\newcolumntype{R}{>{\raggedleft\arraybackslash}X}

\newcommand{\ideasforlater}[1]{}

\acrodef{BH}{black hole}
\acrodef{BBH}{binary black hole}
\acrodef{SN}{supernova}
\acrodefplural{SN}[SNe]{supernovae}
\acrodef{PPISN}{pulsational-pair instability supernova}
\acrodefplural{PPISN}[PPISNe]{pulsational-pair instability supernovae}
\acrodef{NS}{neutron star}
\acrodef{GWTC-4}{the fourth Gravitational-Wave Transient Catalog}
\acrodef{CHE}{chemically homogeneous evolution}
\acrodef{ZAMS}{zero age main sequence}
\acrodef{LVK}{LIGO-Virgo-KAGRA}
\acrodef{IMF}{initial mass function}
\acrodef{AM}{angular momentum}
\acrodef{MS}{main sequence}
\acrodef{LBV}{luminous blue-variable}
\acrodef{SFH}{star-formation history}
\acrodef{CO}{compact object}
\acrodef{DCO}{double compact object}
\acrodef{CE}{common envelope}
\acrodef{CEE}{common envelope evolution}
\acrodef{MT}{mass transfer}
\acrodef{SMT}{stable mass transfer}
\acrodef{CHE}{chemically-homogeneous evolution}
\acrodef{PDF}{probability distribution function}
\acrodef{CDF}{cumulative distribution function}
\acrodef{KDE}{kernel density estimate}

\newcommand{\Msun}{\ensuremath{\xspace M_{\odot}}\xspace}

\newcommand{\Mchirp}{\ensuremath{\xspace \mathcal{M}}\xspace}
\newcommand{\MchirpS}{\ensuremath{\xspace \mathcal{M}_{\mathrm{S}}}\xspace}

\usepackage{tikz,xcolor,hyperref}
\definecolor{lime}{HTML}{A6CE39}
\DeclareRobustCommand{\orcidicon}{\hspace{-1mm}
	\begin{tikzpicture}
	\draw[lime, fill=lime] (0,0)
	circle [radius=0.16]
	node[white] {{\fontfamily{qag}\selectfont \tiny \,ID}};
	\draw[white, fill=white] (-0.0525,0.095)
	circle [radius=0.007];
	\end{tikzpicture}
	\hspace{-3mm}
}
\newcommand{\orcid}[1]{\href{https://orcid.org/#1}{\orcidicon}}
 
\makeatletter
\newcommand{\addres}[1]{\ssmall{#1}}
\newcommand{\email}[1]{\gdef\@email{\url{#1}}}
\makeatother

\begin{document}

\title{ 
New gravitational-wave data support \\ a bimodal black-hole mass distribution\vspace*{4mm}
\\ 
\Large 
Features in the GWTC-4 chirp-mass distribution \\ cannot be explained by traditional remnant-mass prescriptions
} 

\author{
R.\ Willcox$^{1,2}$\thanks{reinhold.willcox@kuleuven.be} ,
F.\ R.\ N.\ Schneider$^{3,4,5}$,
E. Laplace$^{1,2,3,6}$,
Ph. Podsiadlowski$^{7,8,3}$,\smallskip

K. Maltsev$^3$,
I. Mandel$^{9,10}$,
P.\ Marchant$^{11}$,
H. Sana$^{1,2}$,
T. Li$^{1,2}$,
T. Hertog$^{2,12}$
}


\date{August 28, 2025}

\maketitle


\begin{figure}[t!]
    \centering
    \includegraphics[width=0.9\linewidth]{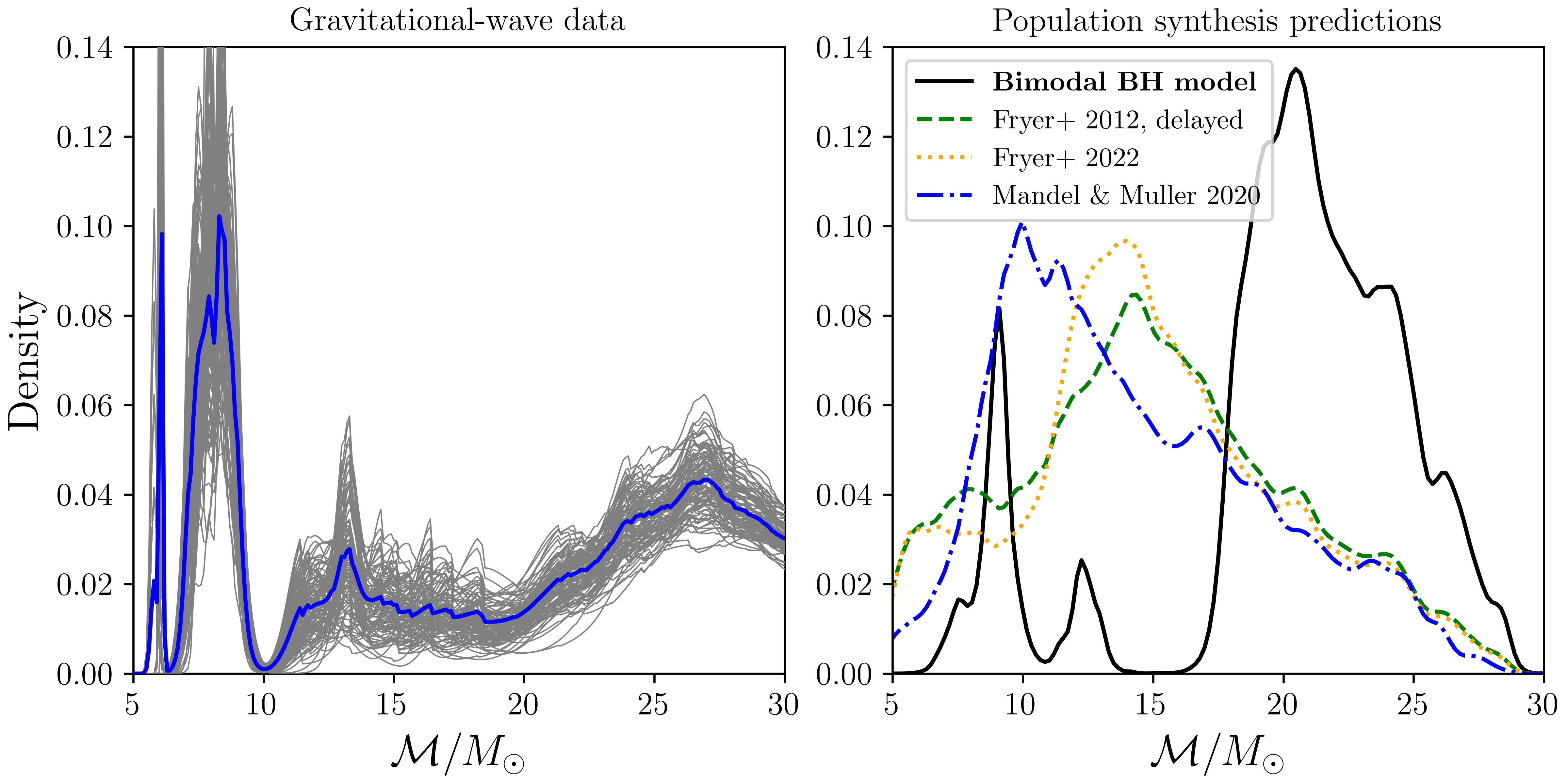}
    \caption{The binary black-hole (BH) chirp-mass distribution from observations (left) and COMPAS population synthesis predictions (right). 
    The chirp mass in both cases refers to the source-frame chirp mass.
    The blue curve on the left is the sum of the posteriors of 158 individual observed events with false alarm rates below 1 per year from all GWTC data releases to date, while the gray curves are obtained by bootstrapping the events to estimate the robustness of features.
    The population synthesis distributions show the predictions for the bimodal black-hole mass model developed in \citet{Schneider_etal.2023_BimodalBlackHole, Maltsev2025} and three commonly used alternatives \citep{Fryer_etal.2012_CompactRemnantMass, Fryer_etal.2022_EffectSupernovaConvection, Mandel_Muller.2020_SimpleRecipesCompact}.
    The predicted distributions are calculated using the cosmic star-formation history from \citet{vanSon_etal.2023_LocationsFeaturesMass} and account for selection effects assuming O3 detector sensitivity.
    Chemically homogeneous stars, which predominantly contribute to higher chirp masses $\gtrsim20~\Msun$, have been excluded here. 
    }
    \label{fig:press_release}
\end{figure}

\begin{figure}[ht!]
\centering
\includegraphics[width=\textwidth]{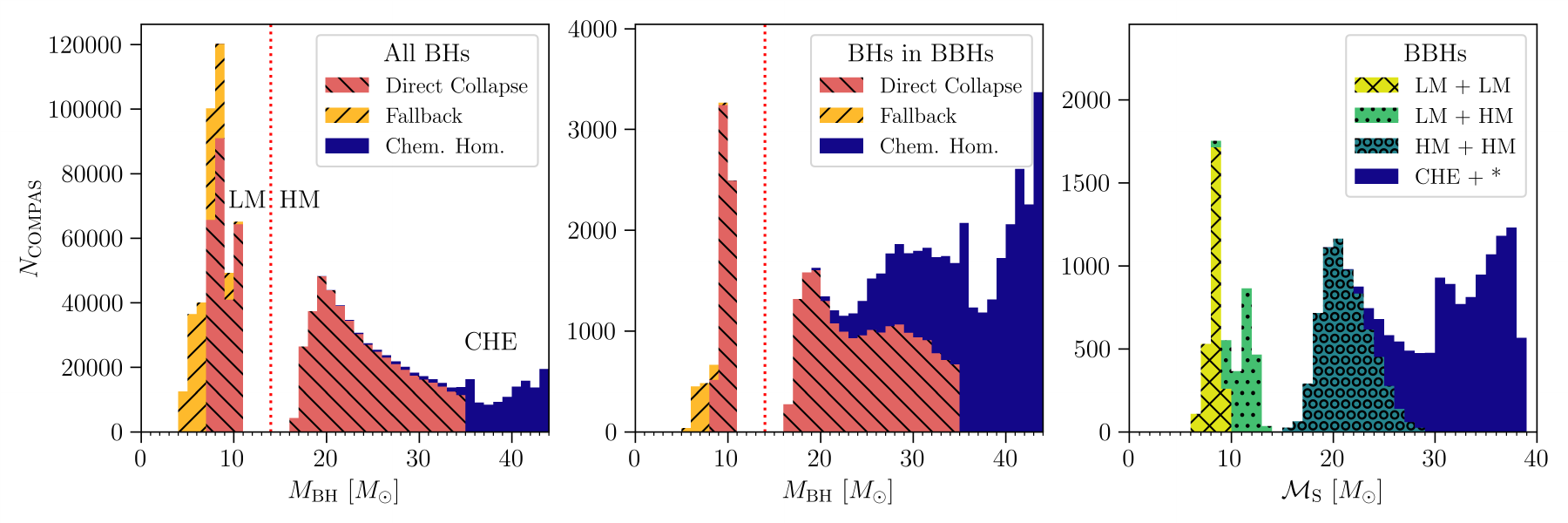}
\caption{
Black-hole (BH) and binary black-hole (BBH) mass distribution from direct collapse, fallback, and chemically homogeneous formation channels, under the bimodal black-hole mass model. 
Left panel: masses for all BHs formed in the simulation. 
Colors and hatching indicate systems that formed via direct collapse (orange) or fallback (yellow), or if they experienced chemically homogeneous evolution (CHE, dark blue), regardless of the explosion mechanism. The vertical dotted line at 14~\Msun indicates the boundary between low-mass and high-mass BHs from the bimodal BH prescription. 
Middle panel: same as the left plot, but including only components of merging BBHs.
Right panel: BBH source-frame chirp masses \MchirpS, prior to convolution with a cosmic star-formation history.
Colors indicate whether both components are low-mass BHs (LM+LM, lime), both are high-mass BHs (HM+HM, sea green), one of each (LM+HM, green), or if either component formed through CHE (dark blue). 
}
\label{fig:mass_spectrum_lmhm}
\end{figure}

{\bf There has been dramatic recent progress in our understanding of the
deaths of massive stars \citep{2007PhR...442...38J, 2020LRCA....6....3M, Heger_etal.2023_BlackHolesEnd, Burrows_etal.2023_TheoryNeutronStar}.  
Detailed stellar evolution studies find a robust, bimodal pattern in the final structures of massive stars \citep{Sukhbold_Woosley.2014_CompactnessPresupernovaStellar,Patton_etal.2022_ComparingCompactObject,Chieffi_Limongi.2020_PresupernovaCoreMassradius,Takahashi_etal.2023_MonotonicityCoresMassive,Temaj_etal.2024_ConvectivecoreOvershootingFinal} whose origin has recently been traced back to their late neutrino-dominated nuclear burning \citep{Laplace_etal.2025_ItsWrittenMassive}. Based on predictions from a simplified neutrino-driven supernova explosion model \citep{Muller_etal.2016_SimpleApproachSupernova}, these patterns can be linked to the formation of black holes.
Applying these to isolated binary stars, \citet{Schneider_etal.2023_BimodalBlackHole} predict a bimodal black-hole mass distribution with a narrow peak around $10~\Msun$ from stars within a narrow range of progenitor properties and a second broader peak starting around $20~\Msun$ from very massive progenitors. 
This bimodal black-hole mass distribution leads to a characteristic distribution of chirp masses $\Mchirp=M_1^{3/5}M_2^{3/5}(M_1+M_2)^{-1/5}$ of merging binary black holes, with two main peaks arising from the merger of two black holes where both come either from the low- or the high-mass peak and a smaller peak in between from the mixed merger of a low-mass and a high-mass black hole \citep{Schneider_etal.2023_BimodalBlackHole}.}\\

The recent Gravitational-Wave Transient Catalog Data Release 4 (GWTC-4, \citealt{gwtc4}) doubles the number of observed compact object mergers with false-alarm rate $< 1~\mathrm{yr}^{-1}$.  This brings to 158 the total sample of confident events with reported chirp-mass measurements ranging between $\sim$1.2 and $\sim$100~\Msun .   We focus here on the range \mbox{$5<\Mchirp/\Msun<30$}, which showed a tentative gap in the chirp-mass distribution around $10~\Msun$ in the previous data release, GWTC-3 \citep{GWTC3}.
Combined with the new data, the observed source-frame chirp-mass distribution now shows a clear peak around $8~\Msun$, a prominent gap at $10~\Msun$ and a rise again up to $\sim$27~\Msun; there may also be a smaller peak around $13~\Msun$ and a dearth between $\sim$15--$20~\Msun$ (see Figure~\ref{fig:press_release}).  We focus our comparison on the chirp mass because in this range it is better measured than individual masses, whose uncertainty prevents robust feature identification \citep{Adamcewicz:2024,GalaudageLamberts:2025}. 

The observed chirp-mass distribution strongly resembles population synthesis predictions utilizing the \emph{Bimodal BH model} (see Figure~\ref{fig:press_release}). These population synthesis predictions are based on a new remnant-mass model \citep{Maltsev2025}, which evaluates several pre-explosion variables obtained from detailed stellar evolution models \citep{Schneider_etal.2021_PresupernovaEvolutionCompactobject,Schneider_etal.2023_BimodalBlackHole,Schneider_etal.2024_PresupernovaEvolutionFinal,Temaj_etal.2024_ConvectivecoreOvershootingFinal} to anticipate the final fate 
of massive stars.  The predictions follow the  \citet{Muller_etal.2016_SimpleApproachSupernova} semi-analytical model and yield a non-monotonic function of carbon-oxygen core mass, metallicity, and mass loss history, including mass transfer onto a binary companion. 

In an upcoming publication (Willcox et al. in prep), we describe the implementation of this bimodal black-hole formation model 
into the rapid population synthesis code COMPAS  \citep{COMPAS:2021,COMPAS:2025} to predict the intrinsic and detectable chirp-mass distributions of merging binary black holes (see Figure~\ref{fig:mass_spectrum_lmhm}).

On the other hand, the predictions from other remnant-mass prescriptions in the literature do not reproduce the peak-gap structure of the gravitational-wave data (see Figure~\ref{fig:press_release}).

The gravitational-wave data show a well defined peak around $8 \Msun$, which is slightly lower than the predicted peak, possibly suggesting slightly more mass loss in the progenitor evolution. The high-mass predicted peak in the bimodal black-hole mass model is more significant than in the data.  This peak is caused by systems that underwent stable mass transfer after the formation of the first black hole. The rate is very sensitive to the details of the treatment of stable mass transfer, in particular, how mass is lost from the system.

If confirmed, a bimodal black-hole mass distribution provides a key observational constraint on the core-collapse supernova mechanism. It may also be used to constrain the evolution of massive stars, as well as uncertain nuclear physics and fundamental binary physics. Meanwhile, robust, redshift-dependent features of the chirp-mass distribution can be utilized as standard sirens to constrain models of cosmological expansion \citep{Schneider_etal.2023_BimodalBlackHole}.

\baselineskip=10pt 

\bibliographystyle{aa}
\bibliography{bib}

\vspace*{6mm} 
\noindent\textbf{\large Author affiliations}\\
\\
\vspace*{1mm}
1. Institute of Astronomy, KU Leuven, Celestijnenlaan 200D, 3001 Leuven, Belgium\\ 
\vspace*{1mm}
2. Leuven Gravity Institute, KU Leuven, Celestijnenlaan 200D, box 2415, 3001 Leuven, Belgium 
\\
\vspace*{1mm}
3. Heidelberger Institut f{\"u}r Theoretische Studien, Schloss-Wolfsbrunnenweg 35, 69118 Heidelberg, Germany  \\
\vspace*{1mm}
4. Astronomisches Rechen-Institut, Zentrum f\"ur Astronomie der Universit\"at Heidelberg, M\"onchhofstr. 12-14, 69120 Heidelberg, Germany \\
\vspace*{1mm}
5. Universit\"at Heidelberg, Department of Physics and Astronomy, Im Neuenheimer Feld 226, 69120 Heidelberg, Germany \\
\vspace*{1mm}
6. Anton Pannekoek Institute of Astronomy, University of Amsterdam, Science Park 904, 1098 XH Amsterdam, The Netherlands \\
\vspace*{1mm}
7. London Centre for Stellar Astrophysics, Vauxhall, London, United Kingdom \\
\vspace*{1mm}
8. University of Oxford, St Edmund Hall, Oxford, OX1 4AR, United Kingdom \\
\vspace*{1mm}
9. School of Physics and Astronomy, Monash University, Clayton, Victoria 3800, Australia\\
\vspace*{1mm}
10. OzGrav, Australian Research Council Centre of Excellence for Gravitational Wave Discovery\\
\vspace*{1mm}
11. Sterrenkundig Observatorium, Universiteit Gent, Krijgslaan 281 S9, 9000 Gent, Belgium \\
\vspace*{1mm}
12. Institute for Theoretical Physics, KU Leuven, Celestijnenlaan 200D, 3001 Leuven, Belgium\\

\end{document}